\documentclass[transmag]{IEEEtran}
\usepackage{latexsym}
\usepackage{graphicx}
\usepackage{amsfonts,amssymb,amsmath}
\usepackage{hyperref}

\def\BibTeX{{\rm B\kern-.05em{\sc i\kern-.025em b}\kern-.08em T\kern-.1667em\lower.7ex\hbox{E}\kern-.125emX}}
\begin{document}

\title{Flexo-extension Analysis of the Neck Using Artificial Vision}

\author{K. Vicente, W. Venegas, C. L. V\'asconez, and I. Zambrano 
\thanks{K. vicente was with Escuela Polit\'ecnica Nacional, Departamento de Mec\'anica, Quito, Ecuador. He is now with Secretar\'ia de Educaci\'on Superior, Ciencia, Tecnolog\'ia e Innovaci\'on, \'Area de Equipamiento EOD-PRETT, Quito, Ecuador (e-mail: vikle52@gmail.com).}
\thanks{W. Venegas and I. Zambrano are with Escuela Polit\'ecnica Nacional, Departamento de Mec\'anica, Quito, Ecuador}
\thanks{C. L. V\'asconez is with Escuela Polit\'ecnica Nacional, Departamento de F\'isica, Quito, Ecuador.}}

\IEEEtitleabstractindextext{\begin{abstract} En el tratamiento del dolor cervical se emplean equipos y t\'ecnicas que no miden la intensidad del dolor del paciente, sino que \'unicamente permiten observar los daños estructurales de dicha regi\'on. Sin embargo, la evaluaci\'on de \'este dolor se puede llevar a cabo al analizar las variables cinem\'aticas de los tres movimientos de la articulaci\'on cervical: flexo-extensi\'on, flexo-lateral y rotaci\'on. En este trabajo estudiaremos la fiabilidad de la t\'ecnica de fotogrametr\'ia, mediante el uso de una c\'amara de bajo costo, denominado Kinect V1. La c\'amara Kinect adquirir\'a los par\'ametros cinem\'aticos del movimiento de flexo-extensi\'on de la articulaci\'on del cuello y, t\'ecnicas de visi\'on artificial y de procesamiento de im\'agenes de profundidad/color del sensor Kinect ser\'an empleadas para obtener las trayectorias de los marcadores anat\'omicos y t\'ecnicos. Se utilizar\'a un filtro de Kalman para corregir el seguimiento continuo de las trayectorias de los marcadores t\'ecnicos y, consecuentemente, las coordenadas espaciales de cada marcador. Los datos fueron obtenidos de siete sujetos de prueba, entre hombres y mujeres, f\'isicamente sanos. Las edades de los sujetos est\'an comprendidas entre 17 y 40 años. Asociados a las coordenadas de cada marcador t\'ecnico, calculamos los par\'ametros cinem\'aticos de velocidad angular, aceleraci\'on angular y desplazamiento angular, para obtener los par\'ametros de fiabilidad y correlaci\'on entre pruebas. Esto \'ultimo se realiz\'o al analizar el error est\'andar medio, el \'indice de correlaci\'on m\'ultiple y los \'indices de correlaci\'on de Pearson, empleados para an\'alisis cl\'inico. El alto \'indice de correlaci\'on entre los ensayos realizados nos permite ratificar la fiabilidad  de nuestra metodolog\'ia. \end{abstract}

\begin{IEEEkeywords}
flexo-extensi\'on, fotogrametr\'ia, Kinect, filtro de Kalman, cinem\'atica
\end{IEEEkeywords}}

\maketitle

\section{Introducci\'on}

\subsection{Antecedentes al estudio del dolor del cuello}
El dolor de cuello (conocido tambi\'en como columna cervical) es un trastorno musculo-esquel\'etico que afecta a la salud o al normal desenvolvimiento de un individuo. En la actualidad, no se conoce sus precursores exactos y tampoco existen datos epidemiol\'ogicos locales \cite{robaina1998cervical, ant2019}. Por ello, las repercusiones a mediano y largo plazo son desconocidas. El sacrificio de la estabilidad del cuello a cambio de la movilidad \cite{magee2013orthopedic}, lo convierte en una zona vulnerable a lesiones, por lo que el estudio de este conjunto de articulaciones es un campo activo. Causas subyacentes, como choques del tejido nervioso, patolog\'ia osteoligamentosa o anormalidades del control neuromuscular \cite{moskovich1988neck,pettersson1997disc,ellingson2013instantaneous}, han sido identificadas como causas del dolor de cuello. A su vez, estas causas pueden haber sido originadas por traumatismos provocados por accidentes, mala postura, infecciones, condiciones inflamatorias, enfermedades reum\'aticas o por enfermedades cong\'enitas \cite{antonaci2000current,de2008clinimetric}. La diversidad de precursores para el dolor del cuello dificulta el diagn\'ostico de la afecci\'on, el mismo que depende fuertemente de la metodolog\'ia empleada para llevarlo a cabo, lo que redunda en un tratamiento inadecuado para tratar la afecci\'on \cite{baydal2013cinematica}. A esta dificultad se le adicionan factores como la estructura anat\'omica, los movimientos compensatorios, el sexo, la edad, la rutina, entre otros \cite{antonaci2000current}.

\subsection{T\'ecnicas y medidas contempor\'aneas}
El comportamiento del cuello se puede cuantificar a trav\'es del rango de movimiento cervical (cervical range of motion), ROM, cuya medici\'on puede ser obtenida usando m\'etodos invasivos o no invasivos. Entre estos \'ultimos se obtienen datos suficientemente confiables con Rayos X \cite{williams2010systematic}, tomograf\'ia computarizada y resonancia magn\'etica \cite{yoganandan2001whiplash}, goni\'ometros, inclin\'ometros \cite{jordan2000assessment,snodgrass2014clinical}, Cybex o an\'alisis cinem\'atico 3D \cite{roozmon1993examining} y esc\'aneres electr\'opticos \cite{antonaci2000current}.\\ La elecci\'on del m\'etodo depende si el objetivo del mismo ser\'a de tipo cribado cl\'inico o con fines investigativos sobre la funcionalidad del cuello. Notamos que el an\'alisis de los movimientos de la columna cervical es m\'as frecuentemente evaluado por motivos cl\'inicos \cite{antonaci2000current}. En este contexto, los m\'etodos m\'as adecuados para una evaluaci\'on diagn\'ostica de la ROM son aquellos basados en  inclin\'ometros, electrogoniometr\'ia y video/fotogrametr\'ia. Sin embargo, estos m\'etodos son poco repetibles al depender de puntos anat\'omicos, los mismos que var\'ian entre pacientes \cite{kraemer1989radio}. Es necesario destacar los estudios realizados por Referencia \cite{cescon2014methodological}, que emplean el sistema de detección \textit{Virtual Reality Rehabilitation System} (VRRS) en Padova, Italia. El equipo VRRS tiene una precisi\'on $\pm 0.2$, y sus sensores son colocados en una diadema, fijados con cinta para evitar tracci\'on. En cuanto a las limitaciones de los equipos basados en inclin\'ometros, destacamos que aunque los movimientos de flexo-extensi\'on y flexi\'on lateral son adecuadamente medidos, las curvas de trayectoria (obtenidas con muestras de movimientos de rotación axial) son del todo no repetibles y, consecuentemente, no fiables \cite{willinger2005modal}.\\

Por su parte, los equipos de video y fotogrametr\'ia garantizan la repetitividad del muestreo, lo que les otorga fiabilidad. En la actualidad, estos sensores son los m\'as empleados para realizar investigaci\'on \cite{baydal2013cinematica}. La t\'ecnica de fotogrametr\'ia reconstruye coordenadas 3D a partir de coordenadas 2D, a trav\'es de la superposici\'on de im\'agenes provenientes de al menos dos c\'amaras. Esto la convierte en una t\'ecnica de medici\'on indirecta. Las im\'agenes 3D son construidas estereosc\'opicamente por triangulaci\'on de puntos hom\'ologos \cite{sanchez2006introduccion}.\\

En general, las t\'ecnicas empleadas basadas en inclin\'ometros o goni\'ometros son poco manejables para los operadores y desagradable para el paciente. Esto repercute negativamente en la  fiabilidad de los datos obtenidos y dificulta una evaluaci\'on precisa. Entonces, es necesario identificar un m\'etodo menos invasivo y cuya operaci\'on sea menos compleja, sin sacrificar su utilidad cl\'inica (e.g. c\'amaras digitales). En este documento determinaremos la curva de flexo-extensi\'on del cuello usando equipos de bajo costo como la c\'amara Kinect de Microsoft. En la Secci\'on \ref{metodos} explicaremos las caracter\'isticas de la c\'amara Kinect, as\'i como el sistema de referencia que usaremos para los marcadores \'opticos. Detallaremos el procesamiento de las im\'agenes de color y profundidad recolectados y el procedimiento para la adquisi\'on de las coordenadas espaciales. Nuestros resultados experimentales de validaci\'on ser\'an resumidos en la Secci\'on \ref{resultados} y nuestras conclusiones se expondr\'an en la Secci\'on \ref{conclusiones}. 

\section{Materiales y m\'etodos}\label{metodos}

\subsection{Sensores \'opticos}
La metodolog\'ia se desarrolla en torno al uso de un sensor Kinect de Microsoft. El dispositivo cuenta con un proyector l\'aser, una c\'amara infrarroja, una c\'amara RGB, micr\'ofonos y un procesador personalizado de la marca $PRIMESENSE PS1080$. Adem\'as, el dispositivo tiene una c\'amara RGB con sensor CMOS (con un filtro de Bayer) con una resoluci\'on de $640 \times 480$ pixels, operando a $30$fps. Finalmente, tiene una c\'amara de profundidad basada en un emisor infrarrojo y una c\'amara infrarroja con sensor CMOS monocrom\'atico cuya resoluci\'on es de $640 \times 480$ pixels, operando a $30$fps. La profundidad de las im\'agenes se determina en funci\'on del tiempo que se demora en reflejar la luz infrarroja \cite{ladinoprograma,nuno2012reconocimiento}. La imagen RGB está compuesta por un conjunto de pixels, compuestos de cuatro componentes: rojo, verde, azul y transparencia (alfa), respectivamente. El \'ultimo canal funciona tambi\'en como RGBa y como vac\'io para imágenes RGB. Cada componente tiene un valor entero de 0 a 254 (correspondiente a byte), por lo que cada pixel posee cuatro bytes. En el caso de la c\'amara de profundidad, cada pixel almacena un valor de intensidad almacenado en dos bytes, lo que provee 2048 niveles de sensibilidad en profundidad. \'Este valor establece la distancia entre el sensor en mil\'imetros \cite{Kinectformatlab2019}.

\subsection{Marcadores t\'ecnicos, anat\'omicos y de referencia}
Para determinar una adecuada toma de muestra establecemos la ubicaci\'on de los marcadores tomando en consideraci\'on resultados previos. La cantidad de marcadores varía en cada metodolog\'ia, a\'un cuando existen ciertas similitudes entre diversos autores. En particular, podemos resaltar el trabajo realizado recientemente usando t\'ecnicas similares de fotogrametr\'ia. Referencia \cite{baydal2013cinematica} expone la utilizaci\'on de un casco con cuatro marcadores reflectantes equidistantes en la cabeza. Cada marcador se encuentra a 20cm del casco con los objetivos de aumentar la inercia del modelo tipo s\'olido rígido y disminuir el error en el c\'alculo de las variables cinem\'aticas. Para establecer los ejes de referencia locales de cada individuo, se coloca un marcador a la altura de la cervical C7 y dos marcadores situados sobre los l\'obulos auditivos. 

Por su parte, Referencia \cite{bertomeu2007nedcerv} usa dos marcadores anat\'omicos colocados en la espalda (a la altura de la cervical C7), dos marcadores anat\'omicos (a la altura de los l\'obulos auditivos) y tres marcadores t\'ecnicos (en la cabeza) empleados durante el movimiento. Referencia \cite{diaz2016dynamic} emplea seis marcadores t\'ecnicos que se encuentran ubicados en la diadema con cuernos, tres en cada cuerno. Para la etapa de referencia emplea marcadores anat\'omicos en los l\'obulos auditivos y dos en el t\'orax. La medici\'on de par\'ametros inerciales se realiza con dos marcadores infraorbitales (ubicados en el punto por debajo de la \'orbita de los ojos) dos puntos tragiones (aleta de la oreja, sobre el trago) y un marcador Sellion, ubicado en el punto de apoyo de las monturas de las gafas sobre la nariz \cite{mcconville1980anthropometric}. 

En el caso de Referencia \cite{grip2007variations} se utiliza un protocolo con 13 marcadores. Cinco marcadores son colocados en la cabeza, en forma de racimo r\'igido, un marcador se coloca en la escotadura supraesternal, tres marcadores en una placa r\'igida en la parte posterior del torso, un marcador en cada hombro y uno en cada fosa mandibular. Los \'ultimos dos marcadores se emplean para estimar la altura cervical y son eliminados antes de la prueba de reposicionamiento. \cite{grip2008cervical} tambi\'en emplean cinco marcadores en forma de racimo en la cabeza. Adem\'as, un marcador en la muesca supraesternal, tres en una placa r\'igida en la parte posterior, una en cada hombro, una en cada fosa mandibular y cuatro en una placa r\'igida en cada brazo superior.  La placa posterior se la coloca a la altura de las v\'ertebras tor\'acicas T6-T8. Lo que totaliza 21 marcadores, analizados con ProReflex del laboratorio Qualisys Medical AB, utilizando cinco c\'amaras infrarrojas para un muestreo a 120Hz.

Referencia \cite{ohberg2003chronic} presentan un tipo de disposici\'on de marcadores diferente: tres marcadores en la cabeza colocadas con unas barras de 11cm y separado a 13cm uno de otro consecutivamente. En el t\'orax del paciente se coloc\'o un chaleco ortopl\'astico con tres marcadores, dos a la altura de la clav\'icula y uno a la altura de la muesca supraesternal.\\

\begin{figure}[h]
    \centering
    \includegraphics[width=0.4\textwidth]{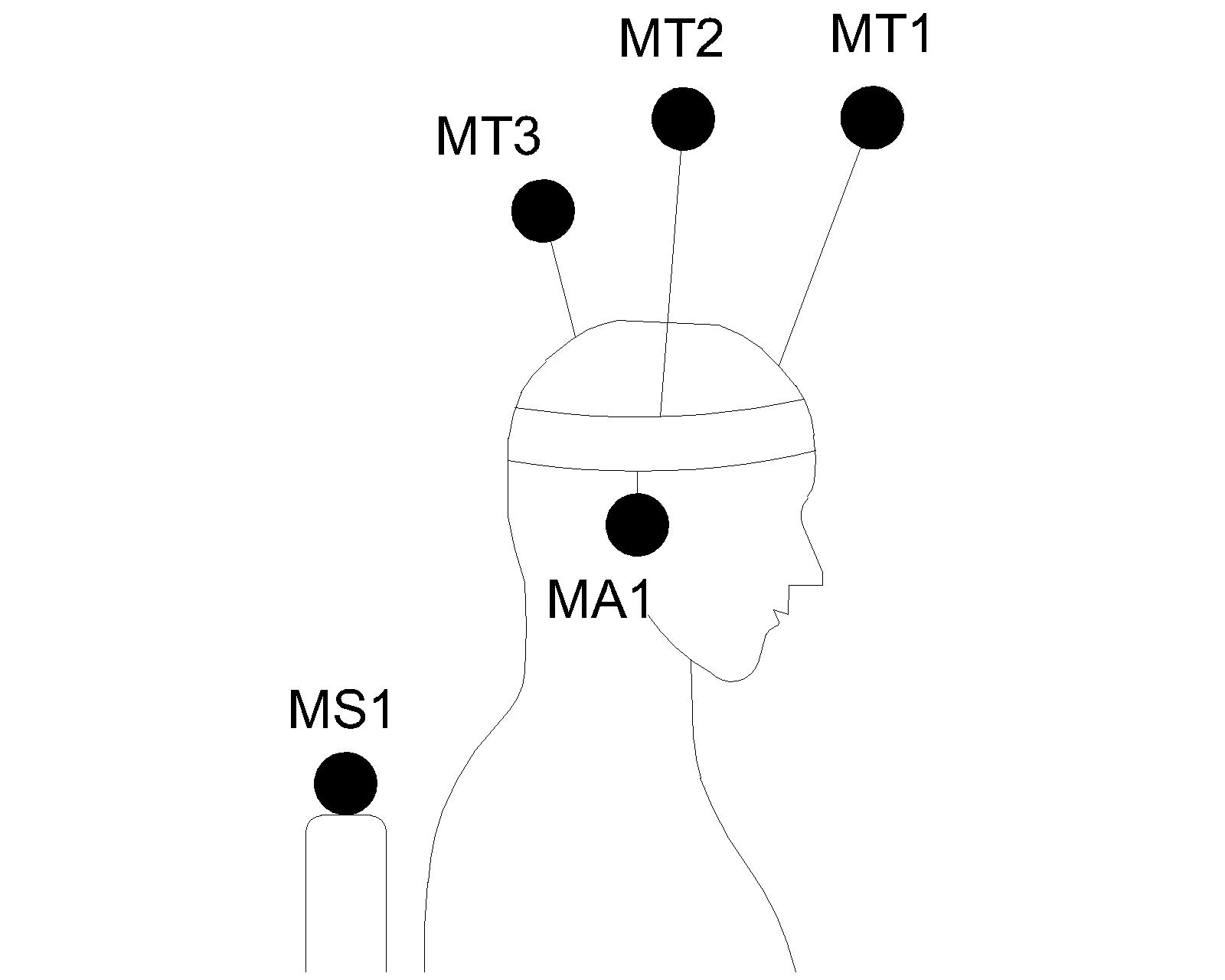}
    \caption{Ubicaci\'on de marcadores en el sujeto de prueba. MT1, MT2 MT3 son los marcadores t\'ecnicos, colocados a 20cm de la cabeza. MA1 es el marcador anat\'omico, ubicado en el l\'obulo auditivo. MS1 es el marcador de referencia de la silla donde se ubica el paciente.}
    \label{marcadores}
\end{figure}

La Figura  \ref{marcadores} muestra la disposici\'on de los marcadores usados en nuestro estudio. La detecci\'on usa cinco marcadores: tres marcadores t\'ecnicos ubicados en una diadema a 20cm de la cabeza, MT1, MT2 y MT3 (con el fin de aumentar la inercia del modelo del s\'olido rígido y disminuir el error en el c\'alculo de las variables cinem\'aticas, e.g., \cite{baydal2013cinematica}). Como marcadores anat\'omicos, colocamos un marcador a la altura del l\'obulo auditivo, MA1, que sirve como un punto de referencia inicial \cite{diaz2016dynamic}. As\'i mismo, usamos un punto de referencia en la silla, MS1, para detectar movimientos de la misma en el muestreo. 

\subsection{Procedimiento para toma de datos}
Hemos establecido que el n\'umero m\'inimo de sujetos de prueba sean 7 y que el n\'umero de muestras por sujeto sean 3, similar al trabajo de Referencia \cite{ruiz2018fiabilidad}. Las personas de prueba var\'ian en peso y talla, y sus edades se encuentran entre 17 a 40 años. Todos los sujetos se encuentran aparentemente sanas, sin ning\'un trauma o enfermedad en las articulaciones del cuello. Se realizan dos muestreos a cada sujeto de prueba en tres sesiones. En el primer caso el paciente no se mueve y se registra las ubicaciones de los cinco marcadores. El segundo muestreo se realiza con movimientos arm\'onicos y naturales de flexo-extensión para registrar la trayectoria de los tres marcadores t\'ecnicos.\\

La c\'amara Kinect se coloca a $\sim 1,90$m desde la pared y a una altura de $\sim 1,05$m. La disposici\'on de la c\'amara cubre totalmente el campo visual donde se ejecutar\'a la prueba. en peso y talla. El centro de coordenadas $(0,0)$ se obtiene de una im\'agen de $640 \times 480$ pixels, localizada en $x_0 = 640/2$ y $z_0 = 480/2$. Gracias al aceler\'ometro del Kinect podemos controlar por software la inclinaci\'on de la c\'amara. De igual manera, verificamos la perpendicularidad de la c\'amara al plano de muestreo al comprobar los valores de profundidad con las cuatro esquinas de la habitaci\'on donde se toma el muestreo.

\subsection{Adquisición de videos RGB y de profundidad}

La obtenci\'on y almacenamiento de las muestras cinem\'aticas del cuello se llev\'o a cabo mediante tres programas. El primer programa almacena simult\'aneamente videos de las im\'agenes RGB y de profundidad. En este programa hemos tomado las siguientes consideraciones:

\begin{itemize}
    \item El video para la imagen RGB es almacenado con formato {\bf Uncompressed AVI}, con extensi\'on ``.avi'' y con una frecuencia de 30fps.
    \item La imagen de profundidad se almacena con formato {\bf Motion JPEG 2000}, con extensi\'on ``.mj2''. y con una frecuencia de 30fps. Debemos considerar que cada pixel tiene 16 bits.
\end{itemize}

\begin{figure}[h]
    \centering
    \includegraphics[width=0.4\textwidth]{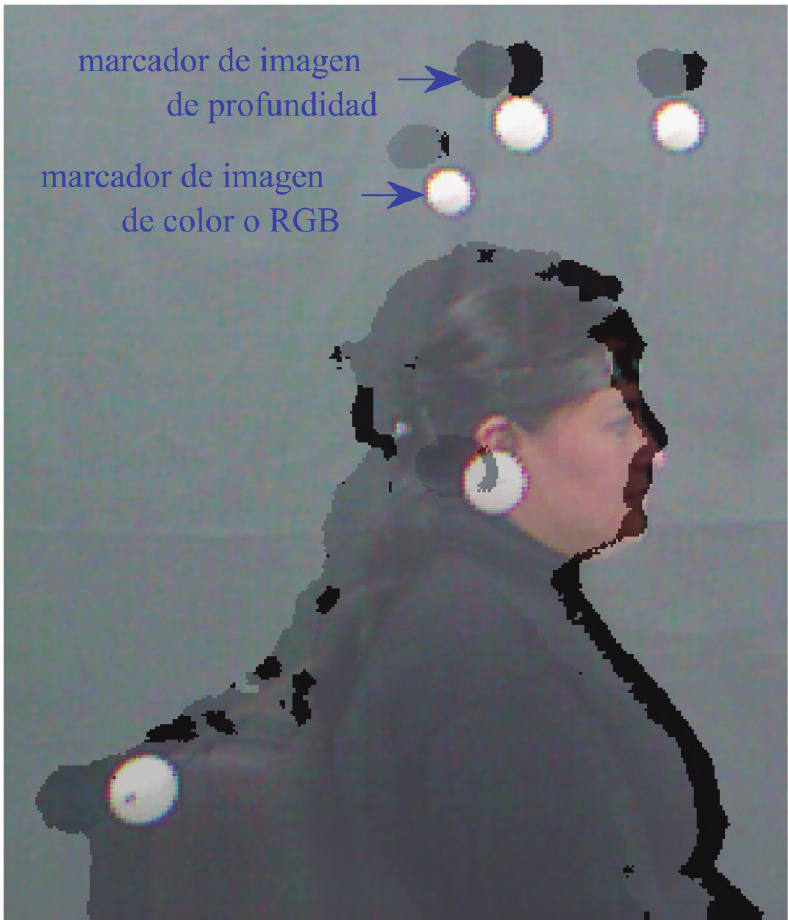}
    \caption{Desplazamiento entre las im\'agenes de color o RGB (c\'irculos blancos) y de profundidad (c\'irculos gris) causado por la ubicación de las c\'amaras en el sensor Kinect.} 
    \label{imagenRGByDEPTH}
\end{figure}

El segundo programa procesa los videos para obtener las trayectorias de los marcadores. El programa realiza un procesamiento con visi\'on artificial de la imagen RGB utilizando herramientas de binarizado y operaciones morfol\'ogicas, presentes en el programa Matlab. Realizamos este procesamiento en cada frame, tanto de la imagen RGB, como de profundidad. En la Figura \ref{imagenRGByDEPTH}
podemos apreciar el desfase entre las dos im\'agenes, propio del arreglo de las c\'amaras internas del sensor Kinect. En particular, de la imagen RGB obtenemos los valores de las componentes $X$ y $Y$, mientras que de la imagen de profundidad obtenemos la componente $Z$.\\

El seguimiento de objetos en movimiento es un tema fundamental dentro de la visi\'on artificial y ampliamente utilizado en muchos campos como la medicina, proyectos militares, v\'ideo vigilancia, navegaci\'on robotizada, entre otros; aunque el seguimiento de m\'ultiples objetos sigue siendo un desaf\'io \cite{park20123d, li2010multiple}.
Para corregir el desplazamiento se utiliz\'o el filtro Kalman, que es un algoritmo que estima el estado del sistema a partir de los datos medidos y luego estimar los estados a partir del error. El filtro permite predecir cu\'al ser\'a la siguiente posici\'on de cada marcador bas\'andose en la posici\'on anterior de cada marcador \cite{patel2013moving}, lo que habilita el seguimiento del mismo. Este procedimiento se basa en 1) la predicci\'on del estado del sistema, que se predice con el modelo din\'amico y en 2) el paso de correcci\'on, que se corrige con el modelo de observaci\'on, de tal modo que el error de la covarianza del estimador es minimizada con un estimador \'optimo \cite{ali2014kalman,Mathworks2018}.\\

Considerando un sistema de seguimiento, donde ${X}_{k}$ es el vector de estado que representa el comportamiento din\'amico del objeto, nuestro objetivo es estimar ${X}_{k}$ a partir de la medici\'on de ${Z}_{k}$, donde el sub\'indice ${k}$ indica el tiempo discreto \cite{li2010multiple}. La descripci\'on matem\'atica se divide en cuatro pasos: 

\begin{enumerate}
    \item \textit{Ecuaci\'on del proceso}  
\begin{equation}\label{uno}
    x_k = \textbf{A}x_{k-1} + w_{k-1},
\end{equation}

donde ${A}$ representa la matriz de transici\'on y ${x}_{k}$ el estado al tiempo ${k-1}$ o ${k}$, El vector ${w}_{k-1}$ es el ruido del proceso, con una distribuci\'on de probabilidad Gaussiana ${p(w)} \sim {N(0,Q)}$.

    \item \textit{Ecuaci\'on de medici\'on}  
\begin{equation}\label{dos}
    z_k = H x_{k} + v_{k},
\end{equation}

donde ${H}$ es la matriz de medici\'on y ${z}_{k}$ es la medida observada al tiempo ${k-1}$ o ${k}$, respectivamente, El vector ${v}_{k}$ es el ruido del proceso,  con una distribuci\'on de probabilidad Gaussiana ${p(v)} \sim {N(0,R)}$.  

\item \textit{Actualizaci\'on de ecuaciones del proceso}  
\begin{equation}\label{tres}
    \hat{x}_{k}^{-} = A \hat{x}_{k-1} + w_{k};
\end{equation}
\begin{equation}\label{cuatro}
    {P}_{k}^{-} = A {P}_{k-1} A^{T} + Q.
\end{equation}

Con el valor de ${z}_{k}$, actualizamos los valores desconocidos de ${x}_{k}$. La estimaci\'on del estado {\it a priori} $\hat{x}_{k}^{-}$ y el error de covariancia $\hat{x}_{k}$ ser\'an obtenidos en el siguiente paso ${k}$.  

    \item \textit{Actualizaci\'on de ecuaciones de medici\'on} 
\end{enumerate}

\begin{equation}\label{cinco}
    K_{k}={P}_{k}^{-} H^{T} (H {P}_{k}^{-} H^{T} + R)^{-1} ;
\end{equation}
\begin{equation}\label{seis}
    \hat{x}_{k} = \hat{x}_{k}^{-} + K_{k}(z_{k}-H \hat{x}_{k}^{-}) ;
\end{equation}
\begin{equation}\label{siete}
    {P}_{k} = (1-K_{k} H) {P}_{k}^{-} .
\end{equation}

El objetivo de este sistema de ecuaciones es la estimaci\'on {\it a posteriori} a trav\'es de ${x}_{k}$, que es una combinaci\'on lineal de la estimaci\'on previa y la nueva medici\'on de ${z}_{k}$. ${K}_{k}$ es la ganancia de Kalman, que se calcula en funci\'on de las ecuaciones de medici\'on actualizadas, despu\'es que la estimaci\'on de estado $\hat{x}_{k}$ {\it a posteriori} y la estimaci\'on de error ${P}_{k}$ {\it a posteriori} son calculadas a partir de ${z}_{k}$. Las ecuaciones de tiempo y de medici\'on se calculan recursivamente con las estimaciones posteriores para predecir una nueva estimaci\'on anterior. Este comportamiento recursivo de estimar los estados es uno de los aspectos más destacados del filtro de Kalman \cite{li2010multiple}.\\

\begin{figure}[h]
    \centering
    \includegraphics[width=0.4\textwidth]{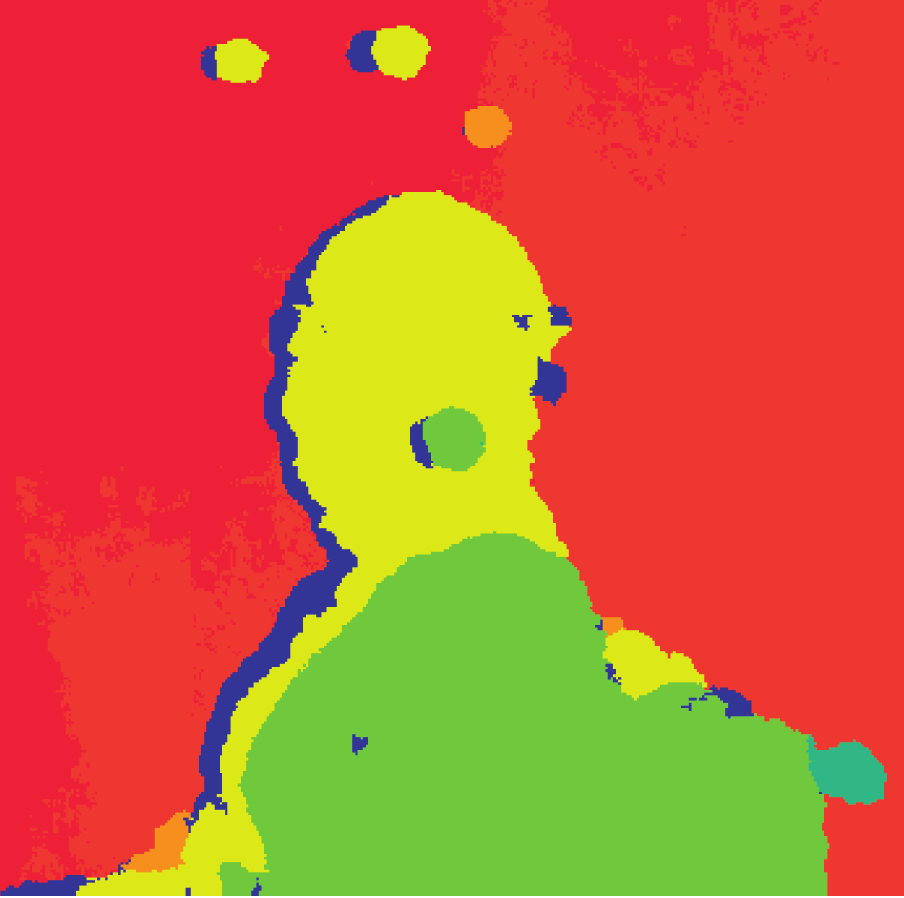}
    \caption{Multi umbrales de intensidad para binarizado de imagen de profundidad. Cada color representa un umbral de intensidad, que nos permite binarizarlo individualmente. Los umbrales de intensidad tienen 11 capas (en promedio). Los tonos azules son la primera capa y representan las lecturas no le\'idas por el sensor Kinect. Los umbrales del 6 al 11 son los mas representativos para la lectura de los marcadores y est\'an en el siguiente orden: cyan, verde, amarillo, naranja, naranja-rojizo y rojo. Los umbrales cyan, verde, amarillo y naranja contienen a los marcadores que nos proveer\'an de sus centroides. Hemos descartado las capas 1, 10 y 11.}
    \label{multiumbrales}
\end{figure}

Para obtener los centroides de los marcadores en la imagen de profundidad, convertimos a la imagen a una escala de grises y luego la binarizamos tomando en cuenta los umbrales promedio de intensidad, tal como se muestra en la Figura \ref{multiumbrales}. Los umbrales de intensidad se obtienen directamente de la imagen adquirida (en $\sim 11$ capas). En esta escala, los tonos azules son la primera capa y representan los errores de lectura del sensor Kinect, debido a los problemas de reflectancia localizados en los bordes. Los umbrales del 6 al 11 son los m\'as representativos para la lectura de los marcadores y est\'an ordenados como: cyan, verde, amarillo, naranja, naranja-rojizo y rojo. Los umbrales del 6 al 9 contienen a los marcadores y son extra\'idos empleando operaciones morfol\'ogicas. Las capas 1, 10 y 11 son descartadas porque pertenecen a las lecturas err\'oneas y al fondo de la imagen, respectivamente. Con la obtenci\'on de los centroides de los marcadores adquirimos en cada punto la componente $Z$ que nos entregar\'a el valor de profundidad expresado en mil\'imetros. Notamos adem\'as que \'esta variable tiene formato {\bf uint16}.\\

\begin{figure}[h!]
    \centering
    \includegraphics[width=0.4\textwidth]{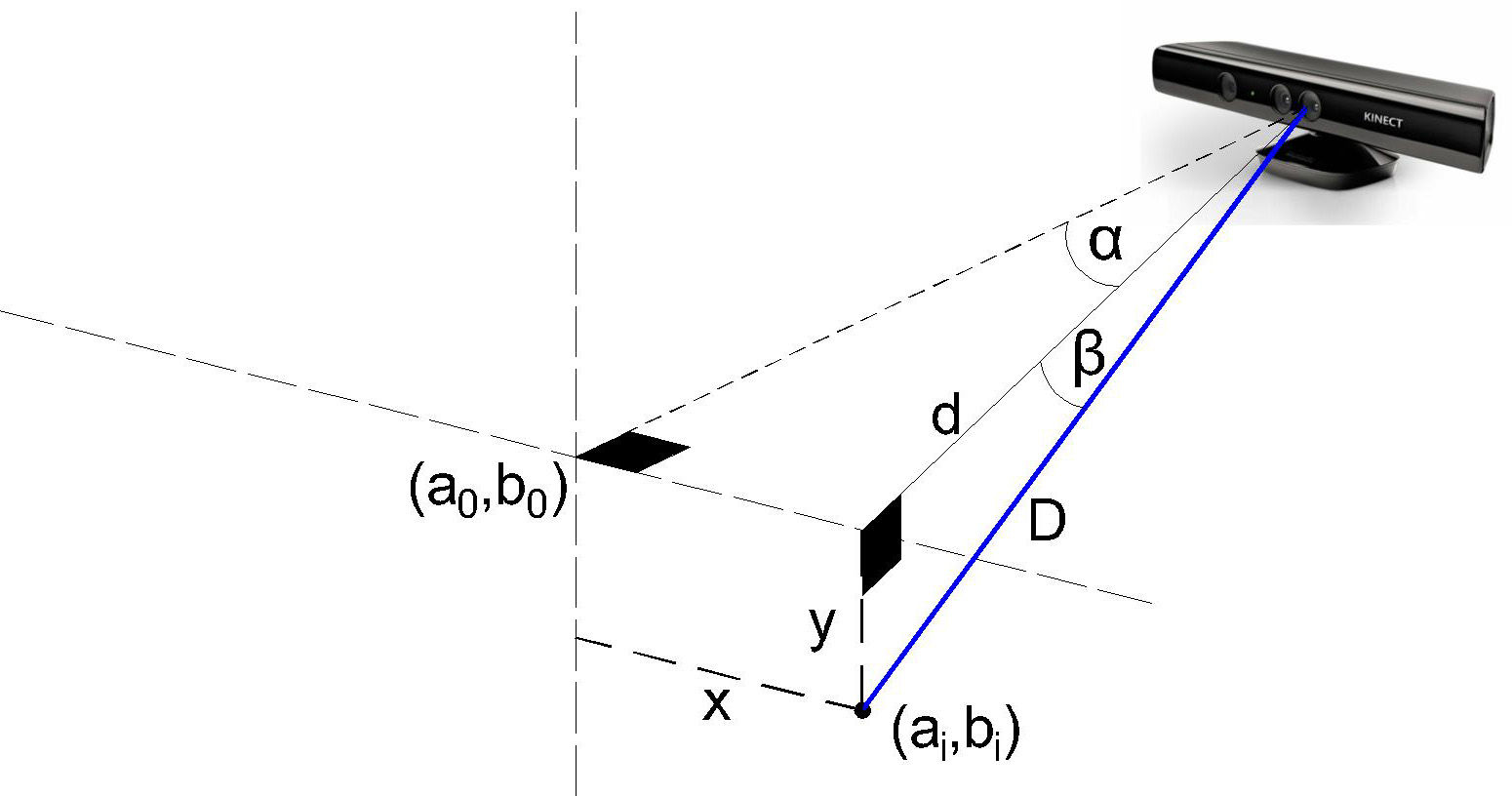}
    \caption{Ubicaci\'on del sensor Kinect respecto al plano que contiene las coordenadas espaciales $X-Y$ de los marcadores. Los \'angulos de apretura $\alpha$ y $\beta$ son utilizados para convertir de pixeles a centímetros cualquier punto coordenado $(a_i,b_i)$, cuyo vector posici\'on es {\bf D}, y su proyecci\'on al eje horizontal ($Y = 0$) es {\bf d}.}
    \label{coordenadaskinect}
\end{figure}

La conversi\'on de unidades de las variables $X$, $Y$ y $Z$ a centímetros se lleva a cabo tomando en consideraci\'on las caracter\'isticas de nuestro sensor \'optico. La cámara Kinect tiene un ángulo de visión de $57^\circ$ en horizontal y $43^\circ$ en vertical, con una resolución de $640 \times 480$ pixels. En nuestro caso de estudio, la disposici\'on geom\'etrica del sensor Kinect se presenta en la Figura \ref{coordenadaskinect}. El Kinect y la imagen obtenida son perpendiculares por lo que establecemos el centro de la imagen como nuestro punto de referencia $(0,0)$, con lo que $a_0 = 320$ y $b_0 = 240$. En el punto $(0,0)$ los ángulos de apertura $\alpha = \beta = 0$ \cite{garciareconocimiento}. De manera general,
\begin{equation}\label{angulobeta}
    \beta =(b_0 - b_i) \frac{43^\circ}{480^\circ \text{pixel}};
\end{equation}

\begin{equation}\label{anguloalpha}
    \alpha =(a_0 - a_i) \frac{57^\circ}{640^\circ \text{pixel}}.
\end{equation}

Por su parte, las variables $X$, $Y$ y $Z$ toman en consideraci\'on la inversi\'on vertical que se aprecia en la misma Figura \ref{coordenadaskinect}.
\begin{eqnarray}\label{varx}
    x &=& -D \sin (\alpha) \cos(\beta);\\
    y &=& D \sin (\beta);\\
    z &=& D,
\end{eqnarray}
\noindent
Donde $D = |${\bf D}$|$.

\section{Resultados experimentales}\label{resultados}

\begin{figure}[h]
    \centering
    \includegraphics[width=0.4\textwidth]{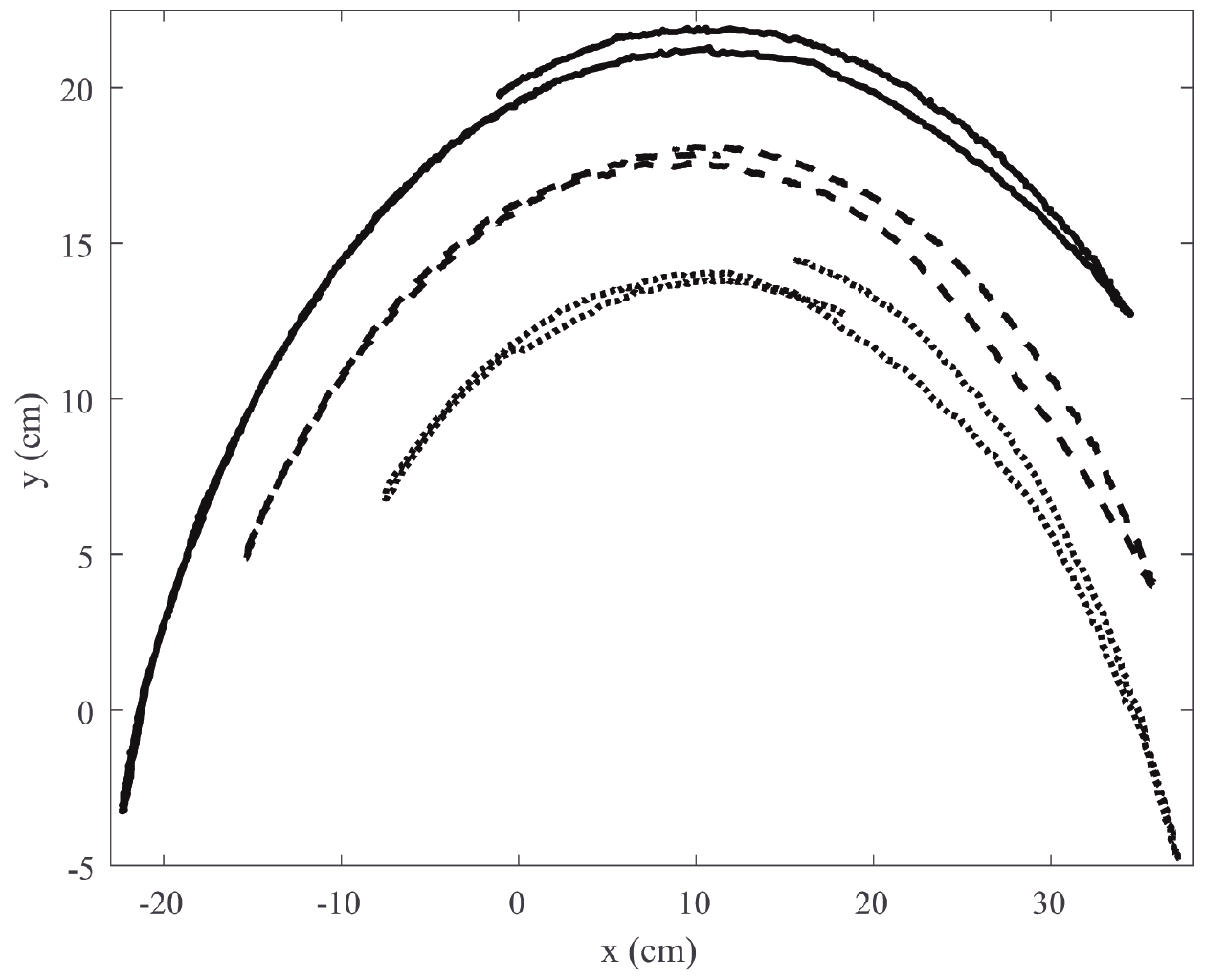}
    \caption{Trayectoria de los tres marcadores t\'ecnicos en el plano $X-Y$. La l\'inea continua representa al marcador MT1, la l\'inea segmentada al marcador MT2 y la l\'inea punteada al marcador MT3.}
    \label{xvsy}
\end{figure}

\begin{figure}[ht]
    \centering
    \includegraphics[width=0.4\textwidth]{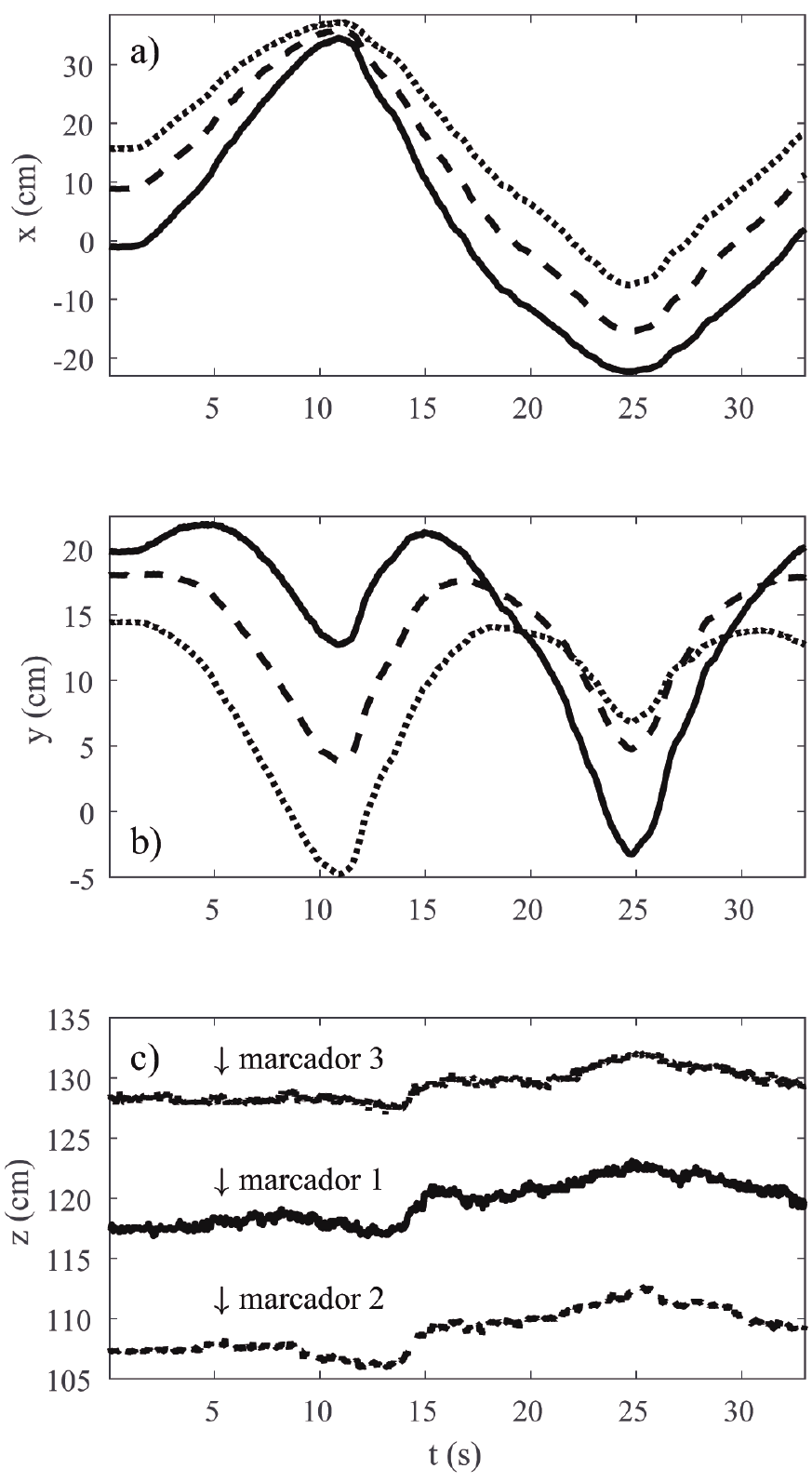}
    \caption{Descomposici\'on $X$ (a), $Y$ (b) y $Z$ (c) de cada trayectoria en funci\'on del tiempo $t$. La l\'inea continua representa al marcador MT1, la l\'inea segmentada al marcador MT2 y la l\'inea punteada al marcador MT3.}
    \label{componentesvst}
\end{figure}

Al observar la Figura \ref{xvsy} apreciamos que las trayectorias de los 3 marcadores t\'ecnicos son continuas y se asemejan al movimiento realizado por los sujetos de prueba. La elecci\'on del filtro de Kalman resulta ser id\'onea. Tal como se muestra en la Figura \ref{componentesvst}, la trayectoria de los marcadores es continua para los 3 componentes y no existe mezcla de datos en el tiempo. Esto podr\'ia ocurrir cuando los marcadores se encuentran en el punto de máxima flexi\'on o en el punto de m\'axima extensi\'on. En los paneles a) y b) se aprecian el punto de m\'axima extensi\'on en $\sim 25$s y el punto de m\'axima flexi\'on en $\sim 11$s, respectivamente, en las componentes $X$ e $Y$. El panel c) evidencia los peque\~nos desplazamientos en la componente $Z$, cuyos valores  pueden interferir en el c\'alculo de las componentes cinem\'aticas.

%\subsection{Validación de parámetros cinemáticos} 
La Tabla \ref{t1} recoge las variables cinem\'aticas m\'as importantes, i.e., \'angulo m\'aximo, rango de movimiento (ROM), velocidad angular media ($\overline{\omega}$) y armon\'ia (definida como la pendiente que forma la aceleraci\'on angular $\dot{\omega}$ con la posici\'on angular $\phi$), obtenidas del an\'alisis de flexo-extensi\'on del cuello para los siete sujetos de prueba y en tres sesiones diferentes.

\begin{table}[h]
\begin{center}
 \caption{Valor medio y desviaci\'on est\'andar (std) de las variables cinem\'aticas del movimiento de flexo-extensi\'on. Los valores fueron obtenidos de los ensayos realizados a todos los sujetos de prueba, en tres sesiones cada uno.}
 \begin{tabular}{l c c}
  \hline
 {\bf Variable} & \textbf{Media} & \textbf{Std} \\ 
 \hline
\'Angulo m\'aximo ($^\circ$) & 37.36 & 17.32 \\
ROM ($^\circ$) & 92.38 & 15.88 \\
$\overline{\omega}$ ($^\circ/$s) & 0.79 & 1.96 \\
Armon\'ia ($^\circ/$s$^2/^\circ$) & 0.46 & 0.09 \\ 
\hline
 \end{tabular}%
 \label{t1}
 \end{center}
 \end{table}

La comparaci\'on directa de las variables cinem\'aticas nos muestra que el valor de armonía es inferior a $0.70$, correspondiente a trayectorias no lineales y poco reproducibles. Los par\'ametros de \'angulo m\'aximo y ROM tienen alta desviaci\'on est\'andar (Std), lo que confirma que, a pesar de existir buen control en el muestreo, existen par\'ametros externos que influyen y afectan al sujeto de prueba.

El an\'alisis de fiabilidad lo realizamos mediante la correlaci\'on de las muestras entre sesiones. En nuestro caso, analizamos la correlación entre $\phi$ (o ROM) respecto a $\overline{\omega}$ y a la posici\'on de referencia concerniente al marcador $MA1$ ($RO$). Observamos que para el caso de la correlación respecto a $\phi$, el error est\'andar medio (SEM) es alto, SEM $\in [10.0$, $18.0]$, pero para la correlaci\'on de $\overline{\omega}$ es menor a $7.25$, lo que se puede considerar como un valor aceptable. Los valores de coeficiente de correlaci\'on m\'ultiple (CMC) y de Pearson son $\sim 1$. Todo esto indicar\'ia que existe un alto \'indice de correlaci\'on entre las diferentes sesiones (tres por cada individuo). Las Tablas \ref{t2}, \ref{t3}, \ref{t4} y \ref{t5} muestran un alto \'indice de correlaci\'on entre las sesiones y los par\'ametros cinem\'aticos de obtenidos de cada sujeto de prueba.\\

\begin{table} [h!tp]
\begin{center}
 \caption{Valores estad\'isticos, SEM, CMC y de Pearson, obtenidos de la correlaci\'on del \'angulo $\phi$.}
\begin{tabular}{l c c c}
\hline
{\bf Variable} & {\bf SEM} & {\bf CMC}  & {\bf Pearson} \\ 
\hline
Sesiones 1 \& 2  & 18.41  & 0.846 & 0.826 \\
Sesiones 1 \& 3  & 14.13  & 0.923 & 0.899 \\ 
Sesiones 2 \& 3  & 10.32  & 0.848 & 0.930 \\ 
\hline
\end{tabular}%
\label{t2}
  \end{center}% el % evita un espacio entre los minipages que haría que nos pasáramos de ancho 
\end{table} 

\begin{table} [h!tp]
\begin{center}
 \caption{Valores estad\'isticos, SEM, CMC y de Pearson, obtenidos de la correlaci\'on de la velocidad angular media $\omega$.}
\begin{tabular}{l c c c}
\hline
{\bf Variable} & {\bf SEM} & {\bf CMC}  & {\bf Pearson} \\
\hline
Sesiones 1 \& 2  & 6.61 & 0.872 & 0.845 \\
Sesiones 1 \& 3  & 7.25 & 0.802 & 0.845 \\ 
Sesiones 2 \& 3  & 4.15 & 0.854 & 0.906 \\ 

\hline
\end{tabular}%
\label{t3}
  \end{center}
\end{table} 

\begin{table} [h!tp]
\begin{center}
 \caption{Valores estad\'isticos, SEM, CMC y de Pearson, obtenidos de la correlaci\'on de RO en funci\'on de la posici\'on cartesiana en $X$.}
\begin{tabular}{l c c c}
\hline
{\bf Variable} & {\bf SEM} & {\bf CMC}  & {\bf Pearson} \\
\hline
Sesiones 1 \& 2  & 2.52 & 0.832 & 0.898  \\
Sesiones 1 \& 3  & 3.17 & 0.859 & 0.815  \\ 
Sesiones 2 \& 3  & 1.51 & 0.857 & 0.908  \\ 

\hline
\end{tabular}%
\label{t4}
  \end{center}
\end{table}

\begin{table} [h!tp]
\begin{center}
 \caption{Valores estad\'isticos, SEM, CMC y de Pearson, obtenidos de la correlaci\'on de RO en funci\'on de la posici\'on cartesiana en $Y$.}
\begin{tabular}{l c c c}
\hline
{\bf Variable} & {\bf SEM} & {\bf CMC}  & {\bf Pearson} \\
\hline
Sesiones 1 \& 2  & 1.85 & 0.821 & 0.803  \\
Sesiones 1 \& 3  & 3.34 & 0.808 & 0.873  \\ 
Sesiones 2 \& 3  & 0.84 & 0.823 & 0.857  \\ 

\hline
\end{tabular}%
\label{t5}
  \end{center}
\end{table} 

\begin{figure}[h]
    \centering
    \includegraphics[width=0.4\textwidth]{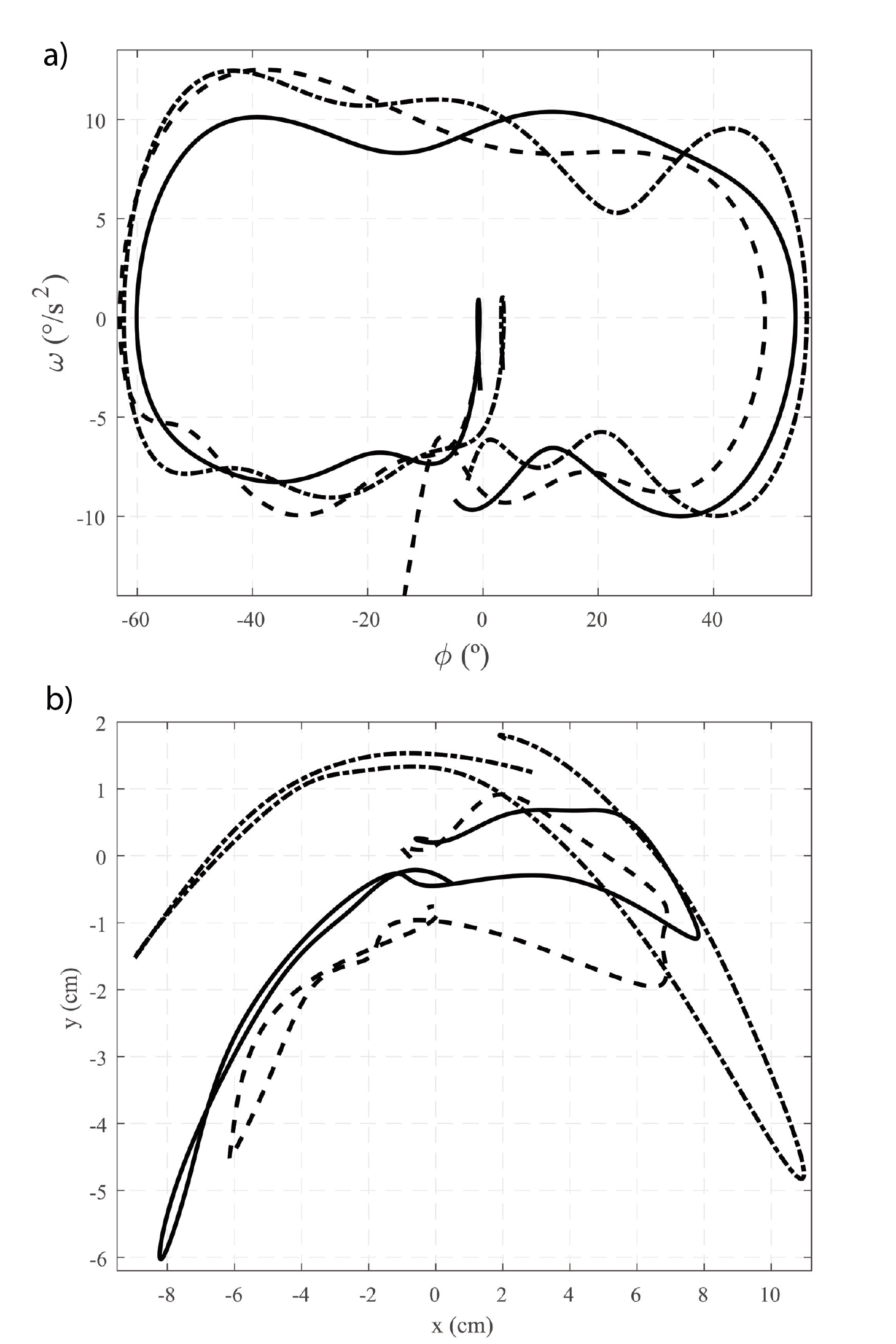}
    \caption{El panel a) se muestra el gr\'afico fasorial de $\omega$ en funci\'on de $\phi$. En el panel b) la curva de desplazamiento suavizada de $x$ en funci\'on de $y$, con centro de referencia en el marcador $MA1$ (ubicado en el l\'obulo auditivo). La l\'inea continua representa la sesi\'on 1, la l\'inea segmentada la sesi\'on 2 y la l\'inea punteada la sesi\'on 3.}
    \label{correlacion}
\end{figure}
Las Figuras \ref{correlacion} muestran la evoluci\'on de la correlaci\'on de $\omega$, en funci\'on de $\phi$ para cada uno de las 3 sesiones (panel a)). Se  evidencia la no-linealidad del sistema por presencia de factores extr\'insecos a las pruebas, y por efecto placebo de los pacientes. En el panel b) observamos la correlaci\'on de las variables $X$ y $Y$ al mover el sistema de referencia hacia el marcador MA1 (l\'obulo auditivo). Hemos suavizado las curvas por cada sesi\'on. Evidenciamos que el SEM de la posici\'on, respecto a la variable $X$ e $Y$, es pequeño ($< 2.70$). Los valores de coeficiente de correlaci\'on m\'ultiple (CMC) y de Pearson son $\sim 1$, lo que indicar\'ia un alto \'indice de reproducibilidad e implica que la correlación múltiple entre todos los ensayos es elevada. Asimismo, el CMC es alto y cercano a la unidad, lo que implica que la correlación múltiple entre todos los ensayos es elevada. Comparativamente, los valores reportados de SEM, CMC y de Pearson son aceptables. Por ejemplo,  Referencia \cite{weir2005quantifying} y Referencia \cite{schwenk2012test} establecen como un SEM aceptable cuando los valores de intraprueba son similares entre sujetos. Asimismo, Referencia \cite{bahat2014neck} y Referencia \cite{atkinson1998statistical} se refieren al coeficiente de correlaci\'on de Pearson como la t\'ecnica m\'as com\'un para evaluar la fiabilidad, y consideran valores aceptables del coeficiente cuando es $> 0.8$. Complementariamente,  Referencia \cite{kadaba1989repeatability} sugiere que CMC $> 0.8$ es un valor aceptable, basado en la repetibilidad de las sesiones entre sujetos. Notamos que los sujetos 5 y 7 alteran los datos obtenidos (no mostrado aqu\'i), pero a su vez nos permiten validar el experimento con un nuevo equipo, con prestaciones aceptables.

\section{Conclusiones}\label{conclusiones}

La medici\'on de los par\'ametros cinem\'aticos del cuello, y su correlaci\'on, es de particular valor en la actualidad. Un m\'etodo que sea confiable y reproducible es necesario.

La técnica presentada en este documento se basa en video y fotogrametr\'ia adquiridos con una cámara Kinect. Si consideramos los valores de SEM ($<  2$), CMC ($> 0.7$) e índice de correlación de Pearson ($> 0.8$) de nuestras medidas cinem\'aticas, podemos concluir que la t\'ecnica es fiable. 

El analizar la ubicación de los marcadores t\'ecnicos, anat\'omicos y de referencia presentada en este trabajo es importante para la reproducción de la técnica. En particular, la ubicación de marcadores anatómicos como en el lóbulo auditivo son necesarios para aquellos estudios dinámicos porque servir\'ia como referencia para el c\'alculo de variables, tales como el eje instant\'aneo de rotaci\'on.

Los datos de profundidad obtenidos con el sensor Kinect, y analizados en la Secci\'on \ref{metodos}, mantienen una buena resolución en el rango de los milímetros, pero se ve afectado por los factores de reflectancia de los objetos cercanos, los que producen datos erróneos. Notamos que es probable que \'este error se haya corregido con el software y/o hardware del Kinect V2. El panel c) de la Figura \ref{componentesvst} evidenci\'o que existen variaciones en el eje $Z$ al momento de adquirir las muestras. Esto nos permiti\'o tomar medidas correctivas para el preprocesamiento de datos.

Al comparar el panel b) de la Figura \ref{correlacion} y la Figura \ref{xvsy}, comprobamos que los datos obtenidos poseen una alta correlaci\'on para cada sujeto de prueba y entre sesiones. Concluimos que nuestra técnica reproduce adecuadamente la curva de trayectoria del cuello para el movimiento de flexo-extensión. Sin embargo, a pesar de \'esta alta correlaci\'on, la no-linealidad se debe corregir mejorando el protocolo de medici\'on y capacitando de una mejor manera a los sujetos de prueba para evitar errores por placebo e involuntarios.

La implementación de los filtros de Kalman fue adecuada para el procesamiento de datos en los algoritmos de visión artificial. Estos filtros evitaron la contabilización de falsos positivos cuando se tomaron datos de imagen binarizada.

Finalmente, resaltamos que nuestra t\'ecnica implementa un sistema de fotogrametría tridimensional fiable y de bajo costo.

\end{document}